\documentclass[10pt,twocolumn,letterpaper]{article}

\usepackage[final]{iccv}      

\definecolor{iccvblue}{rgb}{0.21,0.49,0.74}
\usepackage[pagebackref,breaklinks,colorlinks,allcolors=iccvblue]{hyperref}

\usepackage{amsmath,amsfonts,bm}

\def\eqref#1{equation~\ref{#1}}

\def\1{\bm{1}}

\def\vp{{\bm{p}}}
\def\vq{{\bm{q}}}

\def\vx{{\bm{x}}}

\def\vz{{\bm{z}}}

\DeclareMathAlphabet{\mathsfit}{\encodingdefault}{\sfdefault}{m}{sl}
\SetMathAlphabet{\mathsfit}{bold}{\encodingdefault}{\sfdefault}{bx}{n}

\def\sC{{\mathbb{C}}}

\def\sX{{\mathbb{X}}}

\usepackage{multirow}
\usepackage{booktabs}
\usepackage{algorithm}
\usepackage{algpseudocode}
\usepackage{makecell}

\def\paperID{2204} 
\def\confName{ICCV}
\def\confYear{2025}

\title{StolenLoRA: Exploring LoRA Extraction Attacks via Synthetic Data}

\author{Yixu Wang$^{1,2}$\textsuperscript{$\dagger$}, Yan Teng$^{2}$*, Yingchun Wang$^{2}$, Xingjun Ma$^{1}$\thanks{Corresponding authors:  \\ $\mathtt{<tengyan@pjlab.org.cn, xingjunma@fudan.edu.cn>}$} \\[0.5em]
$^{1}$Fudan University $\quad$
$^{2}$Shanghai Artificial Intelligence Laboratory \\}

\begin{document}
\maketitle
\begingroup\def\thefootnote{\textsuperscript{$\dagger$}}\footnotetext{Work done during internship at Shanghai Artificial Intelligence Laboratory.}\endgroup
\begin{abstract}
Parameter-Efficient Fine-Tuning (PEFT) methods like LoRA have transformed vision model adaptation, enabling the rapid deployment of customized models. 
However, the compactness of LoRA adaptations introduces new safety concerns, particularly their vulnerability to model extraction attacks. 
This paper introduces a new focus of model extraction attacks named \emph{LoRA extraction} that extracts LoRA-adaptive models based on a public pre-trained model. 
We then propose a novel extraction method called \textbf{StolenLoRA} which trains a substitute model to extract the functionality of a LoRA-adapted model using synthetic data. 
StolenLoRA leverages a Large Language Model to craft effective prompts for data generation, and it incorporates a Disagreement-based Semi-supervised Learning (DSL) strategy to maximize information gain from limited queries.
Our experiments demonstrate the effectiveness of StolenLoRA, achieving up to a 96.60\% attack success rate with only 10k queries, even in cross-backbone scenarios where the attacker and victim models utilize different pre-trained backbones. 
These findings reveal the specific vulnerability of LoRA-adapted models to this type of extraction and underscore the urgent need for robust defense mechanisms tailored to PEFT methods.
We also explore a preliminary defense strategy based on diversified LoRA deployments, highlighting its potential to mitigate such attacks.
\end{abstract}    
\section{Introduction}
\label{sec:intro}
Adapting large-scale pre-trained foundation models via Parameter-Efficient Fine-Tuning (PEFT) such as Low-Rank Adaptation (LoRA) \cite{hu2021lora,hayou2024lora+} has become a popular approach for obtaining high-performance models for diverse downstream tasks. 
LoRA enables efficient adaptation of large models to a new task by fine-tuning only low-rank matrices added to specific layers, reducing computational cost and memory usage while preserving the pre-trained model's core knowledge. 
Despite the wide adoption of LoRA, the lightweight and compactness of LoRA parameters raise a new safety concern: \emph{they might be more vulnerable to model extraction (ME) attacks}. 
An ME attack extracts the functionality~\cite{orekondy19knockoff,jagielski2020high,Tramr2016StealingML,liang2024alignment} of a victim model by training a substitute model based on the outputs of the victim model for a certain number of specially designed queries.  
While ME attacks have been extensively studied for traditional models \cite{wang2021black,zhao2024fully,pal2020activethief,zhou2020dast,truong2021data,wang2025honeypotnet}, the vulnerability of LoRA adaptations to ME attacks remains largely unexplored\cite{ma2025safety}. 
This gap is significant, as the public availability of a large number of pre-trained models, coupled with the compactness of LoRA parameters, makes it easier to replicate LoRA-adapted models and compromise intellectual property.



To address this gap, we introduce the concept of \textbf{\emph{LoRA Extraction}}, a novel direction of ME attacks specifically targeting LoRA-adapted models (\emph{i.e.}, ViTs~\cite{dosovitskiy2020image,he2022masked,kirillov2023segment}), based on a publicly available pre-trained model. 
Unlike traditional ME which focuses on replicating the entire model’s functionality, LoRA extraction centers on stealing the \emph{efficient adaptations} encoded within the compact LoRA parameters. 
The adversary’s objective is to reconstruct the victim’s LoRA-adapted model by training their own LoRA-adapted substitute that achieves similar downstream performance using a pre-trained foundation model that is publicly available. 
This attack can manifest in two scenarios: the \emph{identical-backbone scenario}, where the attacker uses the same pre-trained ViT model as the victim, and the more challenging \emph{cross-backbone scenario}, where the attacker uses a different pre-trained ViT.

Existing traditional ME methods largely depend on sample selection from existing datasets or synthetic data generation. 
Sample selection methods \cite{orekondy19knockoff,pal2020activethief,wang2021black,zhao2024fully} typically require searching extensive datasets for in-distribution samples, making them computationally expensive and less practical for LoRA extraction. 
This impracticality arises because 1) inference in LoRA-adapted models is slower due to the large-scale parameters of LoRA-adapted models, and 2) finding appropriate samples is challenging, particularly for domain-specific data fine-tuned with LoRA.
Alternatively, synthetic data generation approaches \cite{zhou2020dast,lin2023quda,kariyappa2020maze,truong2021data} often rely on Generative Adversarial Networks (GANs) \cite{goodfellow2014generative} to produce in-distribution data. 
However, GANs frequently face difficulties in generating high-quality, diverse samples, especially for high-dimensional data. 
As Zhao et al. \cite{zhao2024fully} noted, creating high-dimensional samples (e.g., $224\times224$ images) for effective ME can require millions of queries and is prone to failure due to GAN mode collapse \cite{kossale2022mode,durall2020combating,arjovsky2017towards}. 
These limitations make GAN-based approaches less effective for LoRA extraction, where the tuning often involves high-dimensional image data.

In this work, we propose a novel LoRA extraction method named \textbf{\emph{StolenLoRA}}, which employs synthetic data to train the substitute model. 
To obtain effective in-distribution synthetic data, we leverage Large Language Models (LLMs) to generate diverse textual descriptions based on the target class names and prompt a pre-trained Stable Diffusion model \cite{rombach2022high,sauer2023adversarial} to generate high-quality images.
This allows us to tailor the generated data to the specific downstream task the victim’s LoRA model is fine-tuned for.
Moreover, StolenLoRA introduces a distinctive strategy to improve attack efficiency, \emph{i.e.}, the \textbf{\emph{Disagreement-based Semi-supervised Learning (DSL)}}. 
DSL uses class information from the data synthesis process as pseudo-labels for part of the generated dataset and focuses on queries where the substitute model disagrees with these pseudo-labels.
DSL effectively targets areas of uncertainty, guiding the substitute model’s learning process and refining its alignment with the victim’s behavior. 
To further enhance DSL, we iteratively refine the pseudo-labels based on evolving predictions from the substitute model, improving the quality of synthetic data labels and the effectiveness of the extraction.


In summary, our contributions are as follows:
\begin{itemize}
    \item We explore the vulnerability of LoRA and introduce a novel focus of ME attacks called \emph{LoRA Extraction} which aims to extract the functionality of LoRA-adapted models based on a publicly available pre-trained model.
    
    \item We present \emph{StolenLoRA}, a novel LoRA extraction method that leverages LLM-driven Stable Diffusion to generate high-quality synthetic data, circumventing the need to search large-scale datasets or rely on less dependable GAN-based generation.
    
    \item Through comprehensive experiments on five widely used datasets, we demonstrate the effectiveness of StolenLoRA in both identical- and cross-backbone scenarios, highlighting critical vulnerabilities in LoRA-adapted models. We also explore a preliminary defense mechanism based on diversified LoRA deployments.
\end{itemize}

\section{Related Work}

\textbf{Low-Rank Adaptation (LoRA).} \quad
LoRA~\cite{hu2021lora} is a type of popular Parameter-Efficient Fine-Tuning (PEFT) method~\cite{han2024parameter,houlsby2019parameter} for adapting large pre-trained models for downstream tasks.  LoRA~\cite{hu2021lora} introduces small rank-decomposition matrices as trainable parameters into the pre-trained model. 
The number of trainable parameters is significantly reduced by freezing the original model weights and updating only these lightweight LoRA modules. LoRA inspires numerous variants and extensions, such as Mixture-of-LoRA Experts (X-LoRA)\cite{buehler2024x}, Low-Rank Hadamard Product (LoHa)\cite{hyeon2021fedpara}, and Low-Rank Kronecker Product (LoKr)\cite{yeh2023navigating}, further enhancing its efficiency and flexibility for various applications.
However, the compact nature of these LoRA updates raises concerns about their vulnerability to model extraction attacks. 
This paper investigates the potential for adversaries to extract the efficient adaptations encoded within these compact parameters, specifically focusing on the standard LoRA method~\cite{hu2021lora}.


\noindent \textbf{Model Extraction Attacks.} \quad
Model Extraction (ME) attacks \cite{orekondy19knockoff,pal2020activethief,wang2021black,zhao2024fully,zhou2020dast,lin2023quda,kariyappa2020maze,truong2021data,liang2024alignment} aim to construct a substitute model that mimics the functionality of a victim model by querying its API. 
Existing ME methods can be broadly categorized based on their query generation strategies: \emph{Data-Driven ME} leverage existing datasets \cite{orekondy19knockoff,pal2020activethief,wang2021black,zhao2024fully} to select queries, often employing active learning\cite{ren2021survey} or other search strategies\cite{orekondy19knockoff} to identify informative samples. 
However, this type can incur computational overhead as it necessitates traversing the large-scale dataset to find suitable in-distribution samples.
\emph{Data-Free ME}~\cite{zhou2020dast,lin2023quda,kariyappa2020maze,truong2021data, zhao2024fully} utilize generative models (\emph{e.g.}, GANs\cite{goodfellow2014generative}) to synthesize data for querying the victim model. 
While promising, generating effective queries with high-dimensional inputs can require tens of millions of queries and is prone to failure due to GAN mode collapse~\cite{kossale2022mode,durall2020combating,arjovsky2017towards}.

\begin{figure*}
    \centering
    \includegraphics[width=\linewidth]{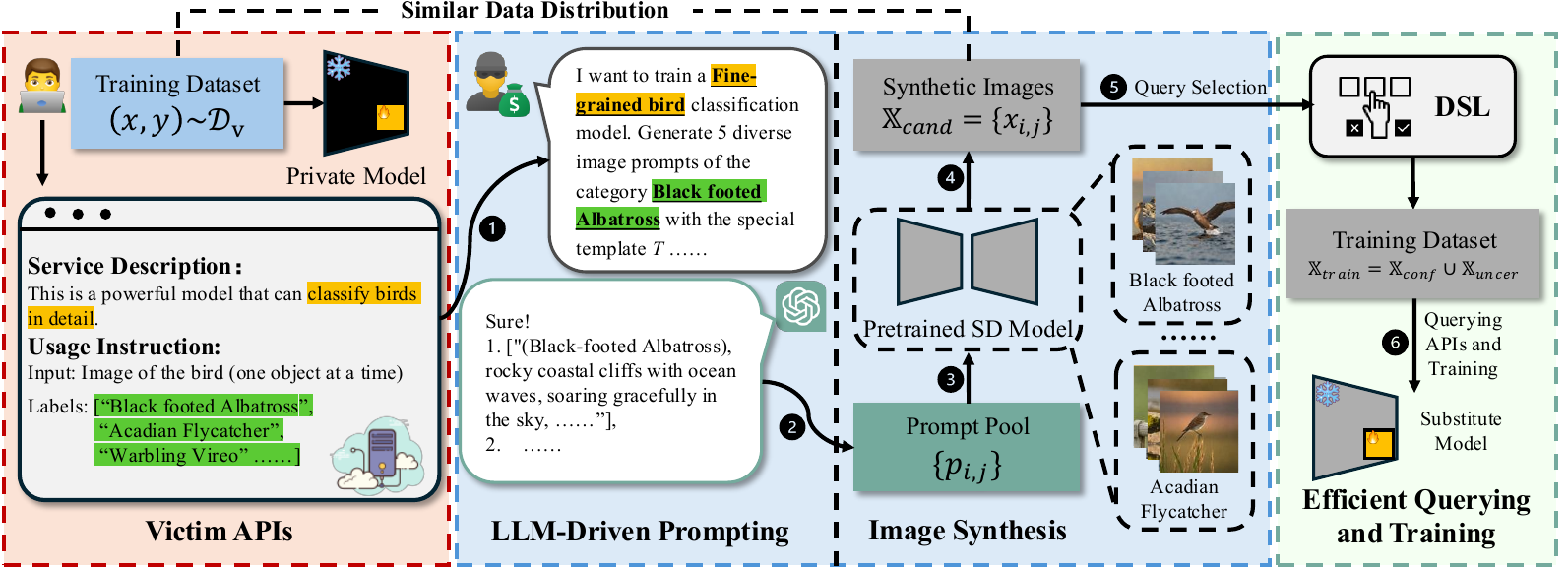}
    \vspace{-0.5cm}
    \caption{Overview of StolenLoRA. It collects target victim model information (dashed red box), synthesizes training data guided by an LLM-driven prompting (dashed blue box), and then efficiently queries the victim API to train a substitute model (dashed green box).}
    \vspace{-0.5cm}
    \label{fig:1}
\end{figure*}

\noindent \textbf{Learning from Synthetic Data.} \quad
The use of synthetic data for training models gains significant traction due to its potential to address data scarcity. 
While traditional synthetic data often lacked the complexity and representativeness of real data, recent advancements in generative models, particularly diffusion models like Stable Diffusion \cite{rombach2022high, sauer2023adversarial}, enable the synthesis of high-quality and diverse images. 
However, scaling the quantity of synthetic data does not necessarily translate to improved performance for training supervised image classifiers. 
As shown by Fan et al.\cite{fan2024scaling}, performance with synthetic data can lag significantly behind training with real data. 
This disparity stems from the limitations of current text-to-image models in accurately generating diverse representations of certain concepts at scale\cite{qin2024diffusiongpt}. 
Efforts to mitigate this gap, such as the distribution-matching framework\cite{yuan2024real}, focus on aligning the distributions of synthetic and real data. 
However, leveraging such high-quality synthetic data for model extraction attacks remains an unexplored area.
\section{Methodology}

\subsection{Problem Formulation}
\label{sec:pf}
\textbf{LoRA Extraction.} \quad
Given a victim model $F \colon [0,1]^d \mapsto \mathbb{R}^{N}$, which is a pre-trained model $F_{base}$ (\emph{e.g.}, ViTs) adapted using LoRA. 
The parameters of $F$\footnote{We omit the parameter for simplicity when there is no ambiguity, \emph{e.g.}, using $F(\vx)$ instead of $F(\vx;\theta)$.} can be represented as $\theta = \{\theta_{base}, \Delta\theta\}$ where $\theta_{base}$ are the parameters of $F_{base}$ and $\Delta\theta$ represents the LoRA updates, which are typically low-rank matrices. 
The objective of a LoRA extraction attack is to train a substitute model $F'$ with parameters $\theta' = \{{\theta'}_{base}, \Delta\theta'\}$ that mimics the functionality of $F$. This is formulated as the following optimization problem:
\begin{equation}
     \arg \min_{\Delta\theta'} \mathbb{E}_{\vx} \mathcal{D}(F(\vx), F'(\vx)),
\end{equation}
where $\vx \in \mathbb{R}^d$ represents an input sample from the victim model task distribution, $\mathcal{D}(\cdot, \cdot)$ represents a measure of functional similarity between the two models.

Depending on whether the attacker has knowledge of $F_{base}$,  we distinguish between two scenarios: \emph{identical-backbone} and \emph{cross-backbone} LoRA extractions.
The substitute model $F'$ can be defined as follows:
\begin{equation}
F'(\vx) =
\begin{cases}
F_{base}(\vx) + \Delta F'(\vx), & \text{identical-backbone} \\
G_{base}(\vx) + \Delta F'(\vx), & \text{cross-backbone}
\end{cases}
\end{equation}
where the former uses the same $F_{base}$, the latter uses different base models $G_{base}$, and the `$+$' symbol denotes the combination of the base model and the LoRA updates.

\noindent \textbf{Distinction from Traditional Extraction.} \quad
Traditional model extraction trains a substitute from scratch, while LoRA extraction leverages a pre-trained base model, inheriting its knowledge and biases. 
Consequently, we hypothesize that a perfectly extracted LoRA substitute might achieve functional equivalence on In-Distribution (ID) data but not on Out-of-Distribution (OOD) data, particularly in the cross-backbone setting, and we experimentally verify this in the experimental section.  
This motivates the use of ID data for effective LoRA extraction. 

\subsection{StolenLoRA}

\noindent \textbf{Overview.} \quad
An overview of our proposed LLM-driven generative attack for extracting LoRA, \emph{StolenLoRA}, is illustrated in Fig.~\ref{fig:1}. 
It comprises two key stages: 1) \emph{Data Synthesis}, where an LLM guides the generation of synthetic data tailored to the victim's task, and 2) \emph{Efficient Querying and Training}, which strategically selects queries and refines labels to maximize information gain from the victim model. 

\subsubsection{Data Synthesis}
The above discussion motivates us to use ID data for extraction. 
However, acquiring such datasets in real-world scenarios is often challenging because the victim model's training data is proprietary. 
While the attacker might not have access to the original training data, they often have access to the model's deployment context, which usually includes a description of the model's functionality and the names of the target classes. 
This information can be leveraged to semantically expand the search space for potential training data. 
Inspired by this, we propose to utilize LLM-driven Stable Diffusion to generate synthetic data based on the class names, effectively bridging the gap between semantic information and visual representations.

\noindent \textbf{LLM-Driven Prompting.} \quad
Given a set of target class names $\sC = \{c_1, c_2, ..., c_n\}$, we employ an LLM to generate detailed prompts for image synthesis. 
We structure the prompts using a predefined template $T$ comprising key visual elements: $T = ``[\text{Subject}, \text{Background}, \text{Angle/Pose}, \text{Lighting}, \text{Style}]"$. 
For each class $c_i$, the LLM generates $m$ variations of the prompt, denoted as $p_{i,j}$, where $j \in \{1, ..., m\}$. This diversification ensures a richer representation of the target class. The prompt generation process is formalized as follows:
\begin{equation}
p_{i,j} = \text{LLM}(c_i, T, \omega_j),
\end{equation}
where $\omega_j$ represents a random seed or instruction to control the variation generated by the LLM.

\noindent \textbf{Image Synthesis.} \quad
We employ a public pre-trained Stable Diffusion model, denoted as $\text{SD}(\cdot)$, to generate images corresponding to the crafted prompts. 
For each prompt $p_{i,j}$, we synthesize one image, resulting in a set of synthetic images $\sX_{i} = \{\vx_{i,j}\}$, where $\vx_{i,j} = \text{SD}(p_{i,j})$. 
The complete synthetic dataset is then $\sX = \bigcup_{i} \sX_{i}$.

\subsubsection{Efficient Querying and Training}
This stage focuses on strategically querying the victim model to refine the substitute model's LoRA parameters while minimizing the number of queries. 
We propose two attack strategies: random learning and DSL.

\noindent \textbf{Random Learning (StolenLoRA-Rand).} \quad
This straightforward approach directly uses the synthesized dataset $\sX$ to query the victim model $F$. 
Then, the substitute model's LoRA parameters $\Delta\theta'$ are trained by minimizing the following objective function:
\begin{equation}
\label{eq:4}
\mathcal{L}'(\Delta\theta') = \mathbb{E}_{\vx \in \sX} \mathcal{L}(F'(\vx), F(\vx)),
\end{equation}
where $\mathcal{L}$ denotes a loss function (\emph{e.g.}, cross-entropy loss). This provides a baseline for evaluating the benefits of more sophisticated attack strategies.

\noindent \textbf{Disagreement-based Semi-supervised Learning (StolenLoRA-DSL).} \quad
While directly using synthesized data for querying and training is feasible, we propose DSL to significantly enhance both the effectiveness and efficiency of the attack.
DSL improves query efficiency by selectively querying the victim model based on the substitute model's uncertainty. Furthermore, it iteratively refines both the pseudo-labels of the synthetic data and the labels obtained from querying the victim, mitigating the impact of noisy pseudo-labels and bridging the gap between the substitute and victim model, especially in cross-backbone scenarios.

\begin{algorithm}[t]
\caption{StolenLoRA-DSL}
\label{alg:StolenLoRA}
\textbf{Input:} Victim model $F$, Class names $\sC$, Query budget $b_t$, Scaling factor $\beta$, Threshold $\tau$, Initial sample size $N$. \\
\textbf{Output:} Substitute LoRA model $F'$.
\begin{algorithmic}[1]
\State $\sX_{0} \gets \bigcup_{i} \text{SynIMG}(\sC, N)$
\State $F'_0 \gets \text{Update}(\sX_{0}, \mathcal{L}_{LR})$ \Comment{{\footnotesize Initial substitute model}}
\State $t \gets 0$, $B_t \gets B$
\While{$B_t > 0$}
    \State $\sX_{cand}^t \gets \text{SynIMG}(\sC, \beta b_t)$ \Comment{{ \footnotesize Generate candidate}}
    \State $\sX_{conf}^t \gets \emptyset$, $\sX_{uncer}^t \gets \emptyset$
    \For{$\vx \in \sX_{cand}^t$}
        \State $\hat{c}, \hat{p} \gets \text{Predict}(F'_t(\vx))$
        \If{$\hat{c} = c(\vx)$ and $\hat{p} \geq \tau$} \Comment{{\footnotesize Disagreement filtering}}
            \State $\sX_{conf}^t \gets \sX_{conf}^t \cup \{(\vx, c(\vx))\}$
        \Else
            \State $\sX_{uncer}^t \gets \sX_{uncer}^t \cup \{\vx\}$
        \EndIf
    \EndFor
    \State $\sX_{query}^t \gets \text{Query}(F, \sX_{uncer}^t, b_t)$ \Comment{{\footnotesize Selective Query}}
    \State $\sX_{train}^t \gets \sX_{conf}^t \cup \sX_{query}^t$
    \State $F'_{t+1} \gets \text{Update}(\sX_{train}^t, \mathcal{L}_{LR})$
    \State $B_t \gets B_t - |\sX_{query}^t|$, $t \gets t + 1$
\EndWhile
\State \Return $F'_t$
\end{algorithmic}
\end{algorithm}

\begin{table*}[t]
    \centering
    \resizebox{\textwidth}{!}{
    \begin{tabular}{l|l|cc|cc|cc|cc|cc}
    \toprule[2pt]
    \multirow{2}{*}{\textbf{Scenario}} & \multirow{2}{*}{\textbf{Method}}  & \multicolumn{2}{c|}{\textbf{CUBS200}} & \multicolumn{2}{c|}{\textbf{Caltech256}} & \multicolumn{2}{c|}{\textbf{Indoor67}} & \multicolumn{2}{c|}{\textbf{Food101}} & \multicolumn{2}{c}{\textbf{Flowers102}}  \\
    \cline{3-12}
    & & Acc & ASR & Acc & ASR & Acc & ASR & Acc & ASR & Acc & ASR \\
    \midrule[1pt]
    \multirow{6}{*}{\makecell[c]{Identical \\ Backbone}} & KnockoffNets & 70.95 & 80.54 & 85.69 & 90.71 & 79.18 & 92.59 & \textbf{82.53} & \textbf{90.62} & 76.24 & 77.28  \\
    & ActiveThief & 72.33 & 82.11 & 86.92 & 92.01 & 77.46 & 90.57 & 80.34 & 88.22 & 77.51 & 78.56  \\
    & DFME & 0.56 & 0.64 & 0.48 & 0.51 & 2.71 & 3.17 & 1.62 & 1.78 & 2.81 & 2.85    \\
    & $E^3$ & 71.94 & 81.67 & 82.36 & 87.18 & 81.43 & 95.22 & 79.65 & 87.46 & 80.67 & 81.77 \\
    & StolenLoRA-Rand & \textbf{75.35} & \textbf{85.54} & \underline{87.62} & \underline{92.75} & \underline{82.38} & \underline{96.33} & 79.00 & 86.75 & \textbf{93.74} & \textbf{95.01} \\
    & StolenLoRA-DSL & \underline{73.23} & \underline{83.13} & \textbf{89.30} & \textbf{94.53} & \textbf{82.61} & \textbf{96.60} & \underline{80.57} & \underline{88.47} & \underline{87.46} & \underline{88.65} \\
    \midrule[1pt]
    \multirow{6}{*}{\makecell[c]{Cross \\ Backbone}} & KnockoffNets & 6.77 & 7.69 & 41.34 & 43.76 & 41.79 & 48.87 & 14.47 & 15.89 & 14.16 & 14.35 \\
    & ActiveThief & 15.42 & 17.50 & 42.89 & 45.40 & 36.04 & 42.14 & 15.11 & 16.59 & 29.09 & 29.49 \\
    & DFME & 0.51 & 0.58 & 0.50 & 0.53 & 1.33 & 1.56 & 0.89 & 0.98 & 1.46 & 1.48  \\
    & $E^3$ & 17.55 & 19.92 & 48.17 & 50.99 & 55.85 & 65.31 & 40.58 & 44.56 & 48.28 & 48.94  \\
    & StolenLoRA-Rand  & \underline{45.70} & \underline{51.88} & \underline{51.75} & \underline{54.78} & \underline{59.18} & \underline{69.20} & \underline{43.16} & \underline{47.39} & \underline{59.29} & \underline{60.10} \\
    & StolenLoRA-DSL & \textbf{50.14} & \textbf{56.92} & \textbf{65.01} & \textbf{68.82} & \textbf{65.07} & \textbf{76.09} & \textbf{44.53} & \textbf{48.90} & \textbf{61.16} & \textbf{61.99}   \\
    \bottomrule[2pt]
    \end{tabular}
    }
    \caption{Effectiveness of StolenLoRA: the Acc (\%) and ASR (\%) of extracted substitute model by different model extraction attacks under 10k queries. (\textbf{Boldface}: the best value, \underline{Underline}: the second-best value.)}
    \vspace{-0.5cm}
    \label{tab:eff}
\end{table*}

The process begins by generating an initial synthetic dataset $\sX^{0} = \bigcup_{i} \sX_{i}^{0}$ and assigning pseudo-labels based on their generating prompts \emph{without querying the victim model}. 
These pseudo-labels, denoted as $c(\vx)$, provide a starting point for training the initial substitute model $F'_0$.
Then, DSL proceeds iteratively, refining both the synthetic data for querying and the substitute model. 
In each iteration $t$, a new set $\sX_{cand}^{t} = \bigcup_{i} \sX_{cand, i}^{t}$ with $\beta * b_t$ samples is generated using the LLM-driven Stable Diffusion process, where $\beta$ is a scaling factor. 
These candidates are then subjected to a \emph{disagreement-based filtering} based on the current substitute model's predictions.
For each candidate $x \in \sX_{cand}^{t}$, the substitute model $F'_t$ predicts a class $\hat{c}$ and associated confidence $\hat{p}$. 
If the predicted class $\hat{c}$ matches the prompt-based pseudo-label $c(\vx)$ and the confidence $\hat{p}$ exceeds a predefined threshold $\tau$, the pseudo-label is considered reliable, and the sample is added to a confidently labeled set $\sX_{conf}^{t}=\{(\vx, c(\vx))\}$. 
Conversely, if the prediction disagrees with the pseudo-label or the confidence is low, the sample is added to an uncertain set $\sX_{uncer}^{t}$.

From the uncertain set $\sX_{uncer}^{t}$, a subset of $b_t$ samples with the lowest confidence scores are selected for querying the victim model $F$. 
These queried samples, along with their true labels obtained from $F$, form the query set $\sX_{query}^{t} = \{(\vx, F(\vx))\}$. 
Combine the confidently pseudo-labeled samples $\sX_{conf}^t$ and the queried samples $\sX_{query}^t$ to form the training set for this iteration: $\sX_{train}^t = \sX_{conf}^t \cup \sX_{query}^t$. 
Crucially, this combined set undergoes \emph{label refining} during training to further address potential inaccuracies in pseudo-labels and mitigate the distribution shift.

Specifically, we employ a label refining strategy.
Let $\vz$ be the logits produced by the substitute model for a given input $\vx$. 
The label refining training loss is calculated as:
\begin{equation}
\mathcal{L}_{LR}(\vz, \vq) = -\sum_{i=1}^C q_i \log\left(\frac{\exp(z_i)}{\sum_{j=1}^C \exp(z_j)}\right),
\end{equation}
where $\vq$ are the soft labels. They are updated using an exponential moving average of the predicted probabilities $\vp$:
\begin{equation}
\vq^{(i+1)} = \mu \vq^{(i)} + (1 - \mu) \vp^{(i+1)},
\end{equation}
where $\mu$ is the momentum parameter. 
This iterative refinement allows the model to learn from both the initial labels and its own evolving predictions, progressively improving the quality of the training data. 
The updated substitute model $F'_{t+1}$ is thus trained on $\sX_{train}^t$ by minimizing:
\begin{equation}
\begin{split}
\mathcal{L}'(\Delta\theta'_{t+1}) &= \mathbb{E}_{\vx \in \sX_{train}^{t}} \mathcal{L}_{LR}(F'_{t+1}(\vx), \vq(\vx)).
\end{split}
\end{equation}

This iterative process continues until the query budget is exhausted, resulting in a final substitute model $F'$ trained on a combination of high-confidence pseudo-labeled data and a smaller set of strategically queried data. 
DSL allows for efficient use of the query budget by focusing on the most informative samples, leading to a more accurate substitute LoRA.
We present the algorithm detail of it in Alg. \ref{alg:StolenLoRA}.

\section{Experiments}
\subsection{Experimental Setup}


\begin{figure*}[ht]
    \centering
    \includegraphics[width=\linewidth]{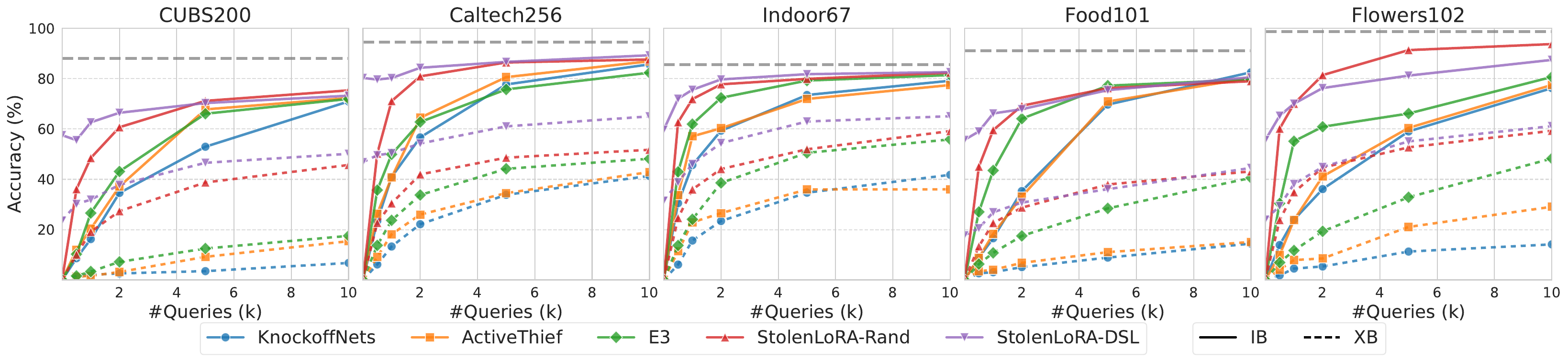}
    \vspace{-0.5cm}
    \caption{Curves of the test accuracy versus the number of queries in IB (solid line) and XB (dashed line) scenarios respectively. The gray dashed line represents the victim models' test accuracy.}
    \vspace{-0.5cm}
    \label{fig:results}
\end{figure*}

\noindent \textbf{Victim Models.} \quad
We use a ViT-Base model pre-trained on ImageNet-21k \cite{ridnik2021imagenet} with additional augmentations and regularization \cite{steiner2021train}.  
This base model is then fine-tuned with LoRA ($r=4$) on five commonly used datasets: CUBS200 \cite{wah2011caltech}, Caltech256 \cite{griffin2007caltech}, Indoor67 \cite{quattoni2009recognizing}, Food101 \cite{bossard2014food}, and Flowers102 \cite{nilsback2008automated}.  
The resulting victim models achieve test accuracies of 88.09\%, 94.47\%, 85.52\%, 91.07\%, and 98.66\%, respectively.

\noindent \textbf{Attack Settings.} \quad
We evaluate LoRA extraction attacks in Identical-Backbone (IB) and Cross-Backbone (XB) scenarios.
In the IB setting, we use the same base model as the substitute base model. 
In the XB setting, we use the ViT-base model pre-trained on ImageNet-1k\cite{russakovsky2015imagenet} with self-supervised masked autoencoder (MAE) method\cite{he2022masked} as the substitute base model.
Both substitute models utilize LoRA with $r=4$ and are trained for 20 epochs using the Adam optimizer\cite{kingma2014adam} with a cosine annealing learning rate schedule\cite{loshchilov2017sgdr} and a base learning rate of 0.01. 

We use GPT-4o mini\cite{gpt4omini} as the LLM in StolenLoRA as it has similar performance at a lower price and can effectively reduce the attack cost.
For the SD model, we use an open-source SDXL-Turbo model\cite{sauer2023adversarial}. 
It only takes 4 sampling steps (compared to SDXL 1.0 which usually requires 40 steps) to synthesize high-quality images, significantly reducing the computational cost.
For the StolenLoRA-DSL method, initial per-category sample size $N$ is set to 10, the scaling factor $\beta$ is 1.5 to not increase too much generation overhead, and the confidence threshold $\tau$ is 0.95. 
Hyperparameter analysis is provided in later sections.

\noindent \textbf{Baselines and Evaluation Metric.} \quad
We compare StolenLoRA against KnockoffNets \cite{orekondy19knockoff}, ActiveThief \cite{pal2020activethief}, DFME \cite{truong2021data}, and $E^3$ \cite{zhu2024efficient} using a query budget of 10k. 
For methods such as KnockoffNets, ActiveThief, and $E^3$ that require real data, we use 3M images in CC3M\cite{sharma2018conceptual} as the attack dataset. 
For $E^3$, which originally selects samples based on semantic similarity between dataset category names and target class names, we improve its effectiveness by calculating semantic similarity between image captions (available in CC3M) and target class names for a more fine-grained sample selection. 
Following prior work \cite{orekondy19knockoff,truong2021data}, we report Test Accuracy (Acc) and Attack Success Rate (ASR), defined as the ratio of the substitute model's accuracy to the victim model's accuracy.

\subsection{Experimental Results}

\noindent \textbf{Effectiveness of StolenLoRA.} \quad
Tab. \ref{tab:eff} presents the effectiveness of StolenLoRA, comparing both the random strategy (StolenLoRA-Rand) and DSL-based (StolenLoRA-DSL) variants against existing extraction attacks. 
Across both identical- and cross-backbone scenarios, StolenLoRA consistently achieves state-of-the-art performance. 
 In the IB setting, StolenLoRA-Rand shows superior performance on CUBS200 (75.35\%) and Flowers102 (93.74\%), while StolenLoRA-DSL excels on Caltech256 (89.30\%) and Indoor67 (82.61\%), maintaining competitive results on other datasets. 
 The XB setting, more challenging due to base model differences, highlights StolenLoRA's significant advantage over baselines, with StolenLoRA-DSL achieving the highest accuracy across all five datasets. 
 This demonstrates DSL's effectiveness in refining synthetic data and improving attack efficiency. 
 Fig. \ref{fig:results} further analyzes query efficiency, showing StolenLoRA-DSL's rapid convergence in the IB setting and consistent progress in the XB setting, outperforming other methods in both scenarios. This efficiency stems from DSL's ability to prioritize informative queries, focusing on uncertain samples, thereby maximizing information gain from each query.

\begin{figure}
    \centering
    \includegraphics[width=\columnwidth]{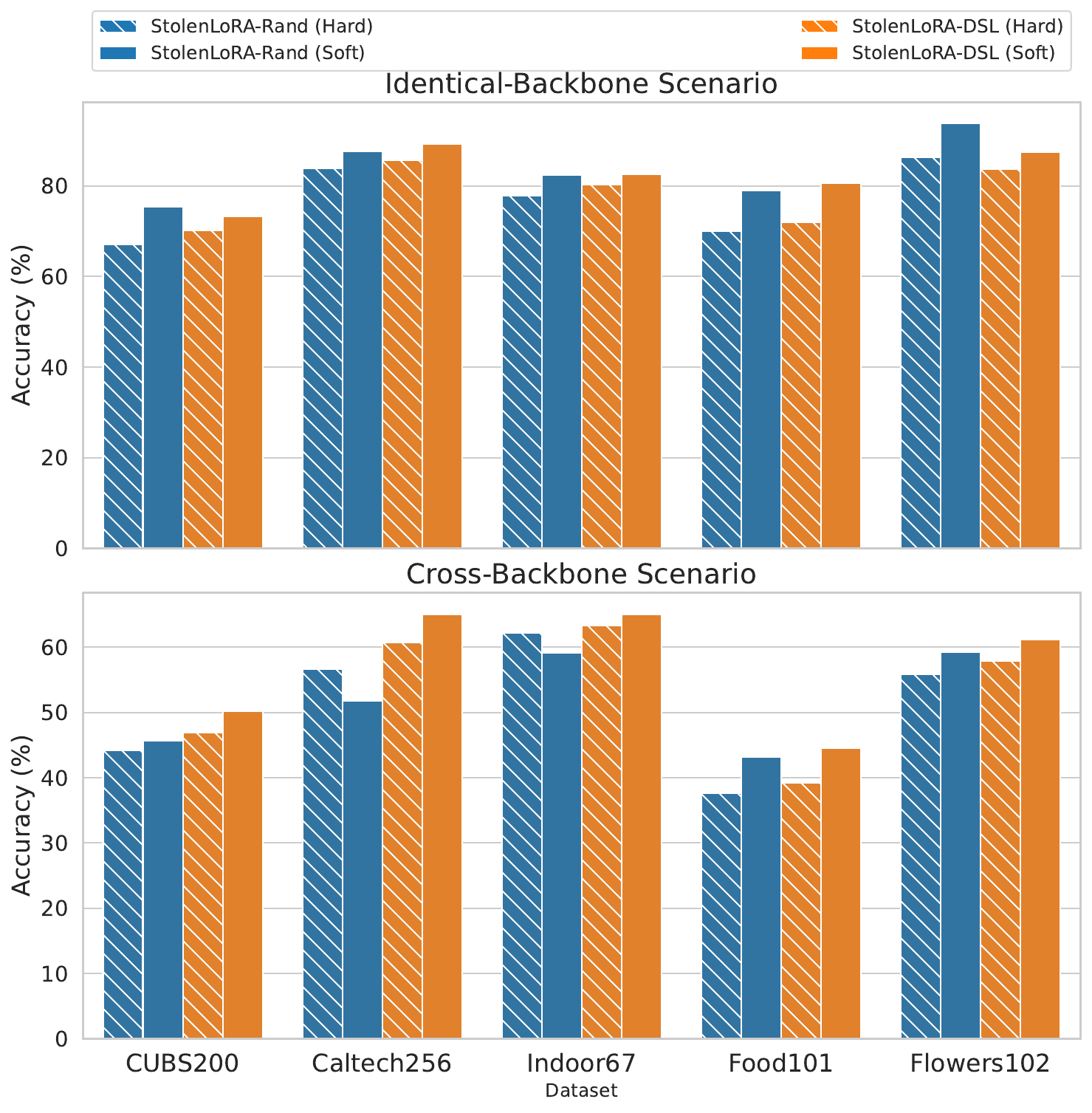}
    \vspace{-0.5cm}
    \caption{Comparison of StolenLoRA performance using hard labels (one-hot) versus soft labels (probabilities) from the victim model.  Results are shown for both Random and DSL in identical- and cross-backbone scenarios across five datasets.}
    \vspace{-0.8cm}
    \label{fig:hard_soft}
\end{figure}

\noindent \textbf{Effectiveness under Hard-Label Scenario.} \quad
We also evaluate StolenLoRA's performance when only hard labels (one-hot encoded predictions) are available, as is the case with some real-world APIs. 
Fig. \ref{fig:hard_soft} presents the results for both StolenLoRA-Rand and StolenLoRA-DSL in the IB and XB settings.
In the IB setting, using hard labels leads to a minor performance decrease. 
For example, StolenLoRA-Rand's accuracy on CUBS200 drops from 75.35\% with soft labels to 67.00\% with hard labels. 
A similar trend is observed for other datasets and StolenLoRA-DSL. 
This suggests that the additional information provided by soft labels is beneficial in the IB setting.
In the XB setting, the impact of hard labels is inconsistent.
While a slight decrease is observed on some datasets, others, like Caltech256 and Indoor67, show a marginal increase in accuracy (from 51.75\% to 56.69\% ). 
This unexpected improvement might be due to the hard labels acting as a regularizer, preventing the substitute from overfitting to the synthetic data, which is more prone to distributional shifts in the XB setting. 
Overall, StolenLoRA exhibits robustness to hard labels in both IB and XB scenarios, demonstrating its practical applicability.

\noindent \textbf{Ablation Study.} \quad
We conduct an ablation study to analyze the contributions of the key components of StolenLoRA: the LLM-driven prompting and the DSL strategy, including the label refining loss.  Tab. \ref{tab:ablation} presents the results on CUBS200 and Indoor67 datasets under both IB and XB settings with a 10k query budget.
Removing the structured prompt template leads to a performance drop in both IB and XB settings, demonstrating the importance of providing structured visual information to the LLM. 
Removing the LLM entirely results in a more significant performance decrease, confirming the crucial role of the LLM in generating effective prompts for diverse and representative synthetic data.  
Furthermore, we evaluate different LLMs, observing that using a more powerful LLM like GPT-4o generally yields better performance than Llama-3.1-8B\cite{dubey2024llama}, particularly in the more challenging XB setting. 
This highlights the benefit of utilizing a stronger LLM for prompt generation.

Disabling DSL and reverting to random querying results in a noticeable performance reduction, especially in the XB setting.  
This underscores the effectiveness of DSL in selectively querying informative samples.  Similarly, removing SAT from the DSL framework also leads to a performance decrease, albeit less pronounced than removing DSL altogether. 
This indicates the benefit of label refining in mitigating the distribution shift between synthetic and real data.

We also analyze the impact of the confidence threshold $\tau$ within DSL in Tab. \ref{tab:tau}.  
The results show that performance generally improves as $\tau$ increases from 0.5 to 0.95.  This indicates that focusing queries on samples where the substitute model is less confident leads to a more efficient use of the query budget.  However, increasing $\tau$ further to 0.99 results in a slight performance decrease, suggesting that an overly stringent threshold can exclude valuable samples from being queried.  We therefore select $\tau=0.95$ as the optimal value for our experiments.

\begin{table}[t]
\centering
\resizebox{\columnwidth}{!}{
\begin{tabular}{l|cc|cc}
\toprule[2pt]
\multirow{2}{*}{\textbf{Method}} & \multicolumn{2}{c|}{\textbf{IB}} & \multicolumn{2}{c}{\textbf{XB}}  \\
\cline{2-5}
& \textbf{CUBS200} & \textbf{Indoor67} & \textbf{CUBS200} & \textbf{Indoor67} \\
\midrule[1pt]
Random & 75.35 & 82.38 & 45.70 & 59.18 \\
- Template & 72.33 & 80.94 & 42.21 & 54.49 \\
- LLM & 70.66 & 74.70 & 39.20 & 47.09 \\
+ Llama-3.1-8B & 73.96 & 81.04 & 46.74 & 58.43  \\
+ GPT-4o & 76.30 & 83.13 & 49.19 & 63.58 \\
\midrule[1pt]
DSL & 73.23 & 82.61 & 50.14 & 65.07  \\
- $\mathcal{L}_{LR}$ & 70.17 & 80.97 & 48.21 & 64.10 \\
\bottomrule[2pt]
\end{tabular}
}
\caption{
An ablation experiment showing the effectiveness of the two modules we designed on CUBS200 and Indoor67 datasets under both IB and XB settings with 10k queries.
}
\vspace{-0.3cm}
\label{tab:ablation}
\end{table}

\begin{table}[t]
\centering
\resizebox{\columnwidth}{!}{
\begin{tabular}{l|cc|cc}
\toprule[2pt]
\multirow{2}{*}{\textbf{$\tau$}} & \multicolumn{2}{c|}{\textbf{IB}} & \multicolumn{2}{c}{\textbf{XB}}  \\
\cline{2-5}
& \textbf{CUBS200} & \textbf{Indoor67} & \textbf{CUBS200} & \textbf{Indoor67} \\
\midrule[1pt]
0.5 & 70.68 & 78.81 & 48.43 &  60.90  \\
0.7 & 71.37 & 80.00 & 48.36 & 63.51  \\
0.9 & \textbf{74.04} & 81.19 & 48.52 & 63.06  \\
\underline{0.95} & 73.23 & \textbf{82.61} & \textbf{50.14} & \textbf{65.07}  \\
0.99 & 73.31 & 81.19 & 49.86 & 61.72  \\
\bottomrule[2pt]
\end{tabular}
}
\caption{
Performance of StolenLoRA-DSL with varying confidence thresholds $\tau$. Results are reported on CUBS200 and Indoor67 under both IB and XB settings with 10k queries.
}
\vspace{-0.5cm}
\label{tab:tau}
\end{table}

\begin{figure}[tb]
    \centering
    \subfloat[Visualization on CUBS200]{
    \includegraphics[width=0.95\columnwidth]{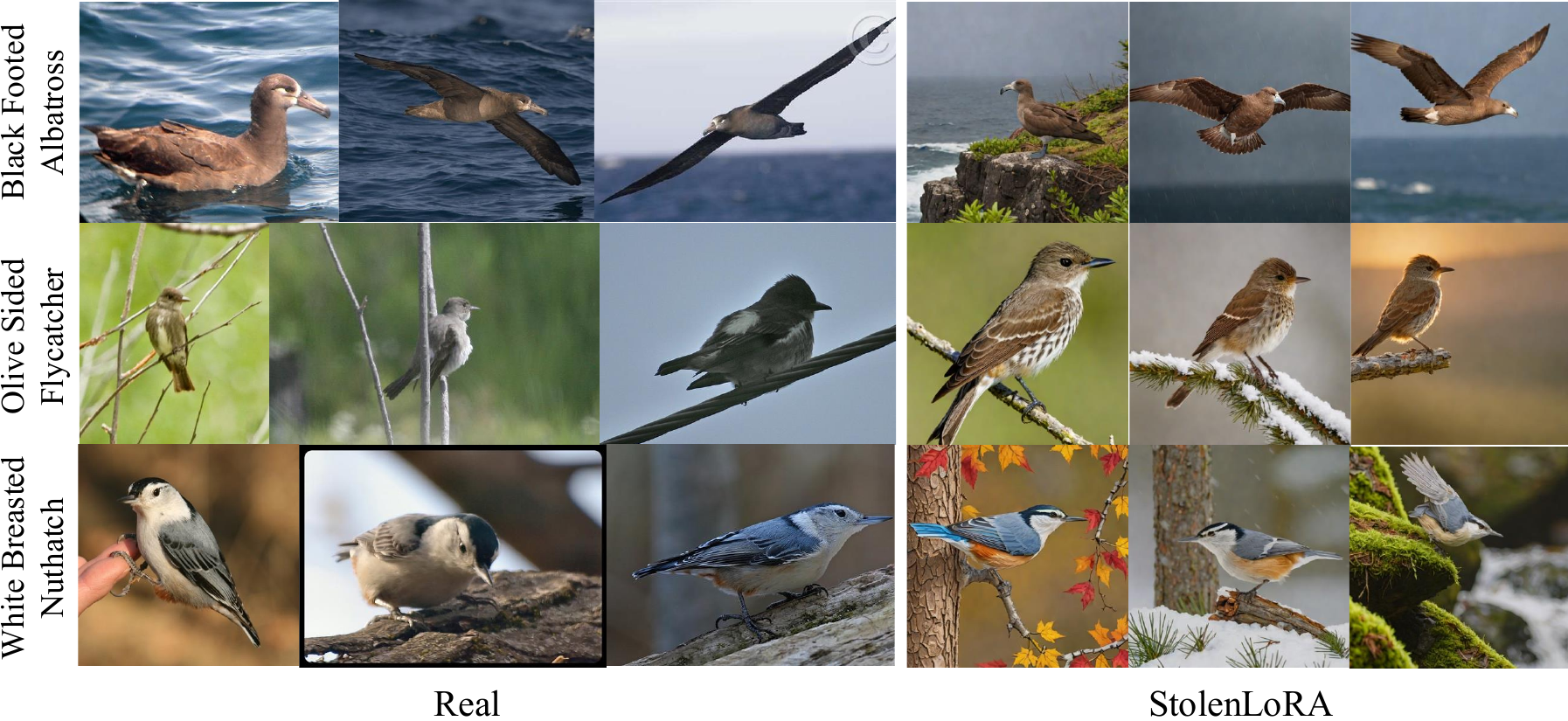}
    }\\
    \subfloat[Visualization on Indoor67]{
    \includegraphics[width=0.95\columnwidth]{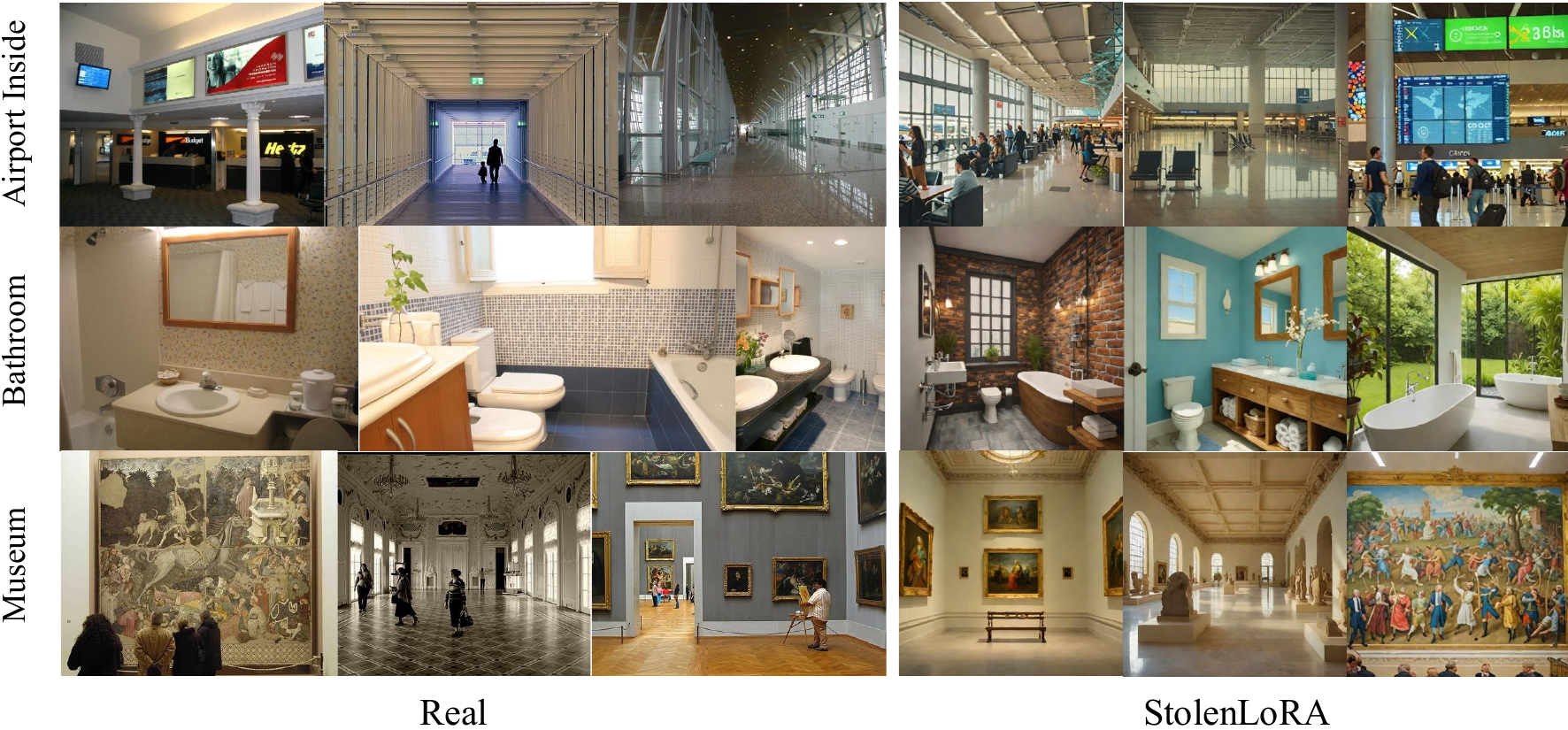}
    }
    \vspace{-0.3cm}
    \caption{ Visualization of StolenLoRA’s synthetic images and corresponding real images of the target category.}
    \vspace{-0.3cm}
    \label{fig:visualization}
\end{figure}

\noindent \textbf{Fidelity and Distribution Analysis of Synthetic Data.} \quad
StolenLoRA's effectiveness relies on the quality of its synthetic training data. 
We evaluate this quality both visually and quantitatively. 
Visually, as shown in Fig. \ref{fig:visualization}, synthetic images generated by StolenLoRA demonstrate a striking resemblance to real images from the target categories, capturing key features and background context crucial for training. 
This visual fidelity is corroborated by significantly lower Fréchet Inception Distance (FID) scores\cite{heusel2017gans} (in Tab. \ref{tab:fid}) compared to other attack methods. 
For example, StolenLoRA achieves an FID of 2.14 compared to 51.44 for KnockoffNets on CUBS200, highlighting a much closer distributional match to the target domain. 
This demonstrates that StolenLoRA's LLM-driven synthesis effectively generates high-fidelity, domain-specific data, even without access to the original training set, which is key to its superior extraction performance.



\begin{table}[t]
    \centering
    \resizebox{\columnwidth}{!}{
    \begin{tabular}{l|ccccc}
    \toprule[2pt]
    \textbf{Method} & \textbf{CUBS200} & \textbf{Caltech256} & \textbf{Indoor67} & \textbf{Food101} & \textbf{Flowers102} \\
    \midrule[1pt]
    KnockoffNets & 51.44 & 8.94 & 22.01 & 50.04 & 55.77 \\
    ActiveThief & 49.33 & 6.01 & 15.72 & 47.84 & 50.19 \\
    DFME & 220.53 & 179.45 & 225.84 & 188.29 & 204.79 \\
    $E^3$ & 41.57 & \textbf{4.29} & \textbf{6.68} & 25.51 & 34.82 \\
    StolenLoRA & \textbf{2.14} & 5.59 & 6.93 & \textbf{19.54} & \textbf{16.63}\\
    \bottomrule[2pt]
    \end{tabular}
    }
    \vspace{-0.3cm}
    \caption{FID scores ($\downarrow$) between the datasets used by each attack method and the victim model's training data distribution.}
    \vspace{-0.5cm}
    \label{tab:fid}
\end{table}

\begin{figure}
    \centering
    \includegraphics[width=\columnwidth]{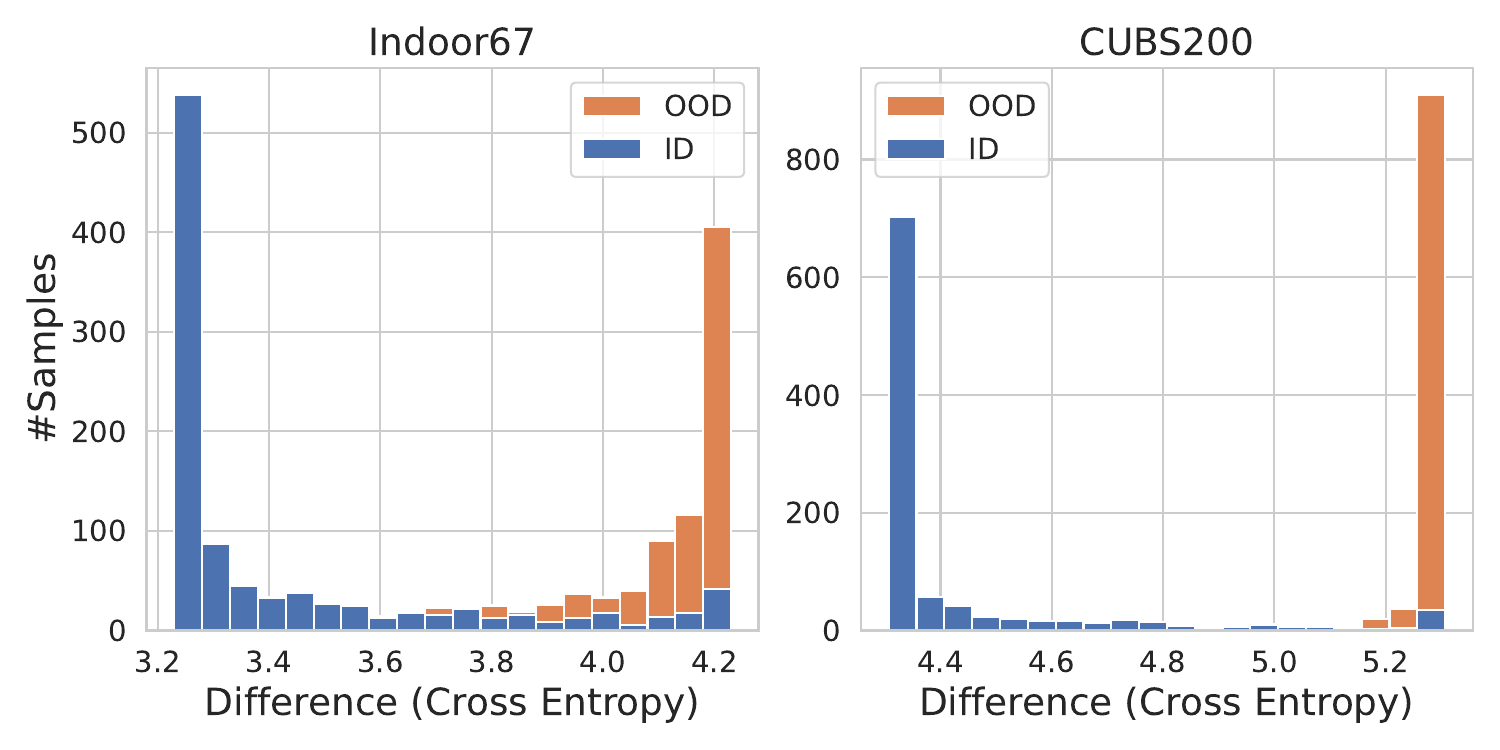}
    \caption{Cross-entropy difference between the victim model and a near-perfectly substitute model on In-Distribution (ID) and Out-of-Distribution (OOD) data. The substitute model is trained using the cross-backbone setting on the victim's training dataset. }
    \vspace{-0.5cm}
    \label{fig:dis}
\end{figure}

\noindent \textbf{Verifying the Distinction Hypothesis.} \quad
As mentioned in Sec. \ref{sec:pf}, we hypothesize that the performance of a perfectly extracted surrogate model will be consistent across In-Distribution (ID) data with the victim model, but not necessarily Out-of-Distribution (OOD) data, motivating us to perform extraction attacks using ID data.
To empirically verify this conjecture, we train a substitute model using the XB setting on the victim model's training dataset, aiming to achieve a near-perfect extraction of the LoRA adaptations. 
We then evaluate both the victim model and the extracted substitute on two sets of data: ID data, represented by the victim model's held-out test set, and OOD data, sampled from the CC3M dataset. 
For each sample, we calculate the cross-entropy between the victim model's predictions and the substitute model's predictions, providing a measure of their functional divergence.
Fig. \ref{fig:dis} presents the distribution of these differences.
On both CUBS200 and Indoor67, the difference is smaller for the ID data compared to the OOD data. 
This observation supports our conjecture that even near-perfect LoRA extraction yields a substitute model that exhibits functional equivalence primarily within the ID domain. 
The divergence in OOD data likely stems from the inherent differences between the pre-trained backbone used by the victim and the substitute. 
This result underscores the importance of utilizing ID data for effective LoRA extraction, as the attacker primarily seeks to replicate the victim's specialized adaptations rather than the general capabilities.

\section{Defending Against LoRA Extraction}
While this paper primarily focuses on the feasibility of LoRA extraction, we also explore a potential defense mechanism based on deploying multiple, diverse LoRA adapters to increase attacker uncertainty.

\noindent \textbf{Dual LoRA with Diversified Predictions.} \quad
Our proposed defense involves training two LoRA adapters, denoted as $\mathrm{L}_A$ and $\mathrm{L}_B$, to maximize the difference in their output distributions while maintaining comparable performance on the target task.  
Instead of minimizing accuracy as initially conceived, we found that maintaining high accuracy while maximizing divergence is crucial for practical deployment. 
We achieve this by optimizing the following loss function:
\begin{equation}
    \mathcal{L}' = \mathcal{L}(\mathrm{L}_A) + \mathcal{L}(\mathrm{L}_B) - \lambda \times \mathrm{KL}(\mathrm{L}_A || \mathrm{L}_B)
\end{equation}
where $\mathcal{L}$ represents the cross-entropy loss for the target task, KL represents the Kullback-Leibler divergence between the output distributions of the two LoRAs, and $\lambda$ is a hyperparameter controlling the balance between task performance and divergence.
At deployment, we randomly select either $\mathrm{L}_A$ or $\mathrm{L}_B$ for each incoming query to generate the prediction. 
This random selection introduces uncertainty and makes it challenging for the attacker to extract the underlying LoRA function consistently.

\begin{figure}
    \centering
    \includegraphics[width=\columnwidth]{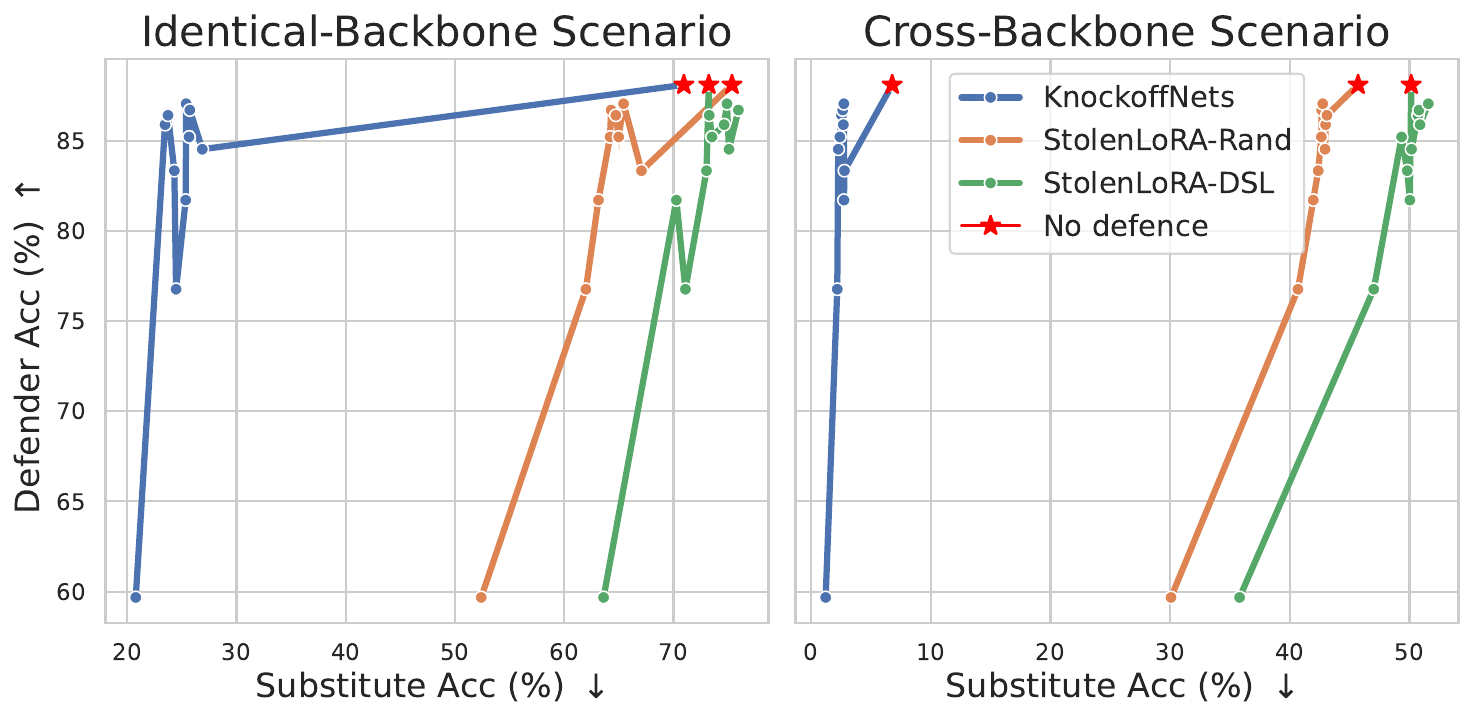}
    \caption{Defneder Acc versus Substitute model Acc trade-off for defenses evaluated against StolenLoRA and KnockoffNets on the CUBS200 dataset with 10k queries.}
    \vspace{-0.8cm}
    \label{fig:defend}
\end{figure}

\noindent \textbf{Experimental Evaluation.} \quad
We evaluate the proposed defense mechanism under both IB and XB settings with 10k queries. 
Fig. \ref{fig:defend} illustrates the trade-off between defender Acc and substitute model Acc. 
The defense consistently degrades the substitute model's performance. 
KnockoffNets experiences substantial accuracy drops (from 70.90\% to 20.78\% in IB and from 6.77\% to 1.21\% in XB), and StolenLoRA-Rand shows similar vulnerability (decreasing 22.95\% in IB and 15.63\% in XB). 
However, StolenLoRA-DSL demonstrates greater resilience, with comparatively smaller accuracy reductions (9.63\% in IB and 14.34\% in XB). 
This resilience is attributed to DSL's efficient query selection and label refining, focusing on informative samples and mitigating the uncertainty introduced by the defense. 
While this defense effectively hinders less sophisticated attacks, it underscores the need for more robust mechanisms to counter advanced techniques like DSL that can learn effectively from limited queries.

\section{Conclusion}
This paper introduces \emph{LoRA Extraction}, a novel model extraction attack targeting the widespread practice of adapting large vision models with LoRA. 
We present \emph{StolenLoRA}, a highly effective method leveraging LLM-driven Stable Diffusion to generate task-specific, high-fidelity synthetic training data, bypassing the limitations of traditional extraction techniques reliant on large-scale real datasets or unreliable GANs. 
Furthermore, StolenLoRA incorporates \emph{Disagreement-based Semi-supervised Learning (DSL)} to efficiently query the victim model, focusing on uncertain predictions and iteratively refining labels, enabling successful extraction with limited queries. 
Extensive experiments demonstrate StolenLoRA's strong performance across diverse datasets and in challenging cross-backbone scenarios. 
We also explore a preliminary defense based on diversified LoRA deployments, showing promise in mitigating these attacks. This work represents a crucial first step towards understanding and addressing the safety risks associated with LoRA adaptation, paving the way for more secure deployments of efficient fine-tuning methods.

\section*{Acknowledgments}
This work is in part supported by National Key R\&D Program of China (Grant No. 2022ZD0160103) and National Natural Science Foundation of China (Grant No. 62276067), and Shanghai Artificial Intelligence Laboratory.
{
    \small
    \bibliographystyle{ieeenat_fullname}
    \bibliography{main}
}

\end{document}